\documentstyle[graphicx,multicol,prl,aps]{revtex}

\newcommand\gsim{\raisebox{-2pt}{$\stackrel{>}{\scriptstyle{\sim}}$}}

\begin{document}


\title{Evidences Against Temperature Chaos in Mean Field and 
Realistic Spin Glasses}

\author{Alain Billoire}

\address{
  CEA/Saclay,
  Service de Physique Th\'eorique,
  91191 Gif-sur-Yvette, France.
}

\author{Enzo Marinari}

\address{
  Dipartimento di Fisica and INFN, Universit\`a di Roma {\em La Sapienza},\\
  P. A. Moro 2, 00185 Roma, Italy. }

\date{\today}                                                 

\maketitle

\begin{abstract}  
We discuss temperature chaos in mean field and realistic $3D$ spin
glasses. Our numerical simulations show no trace of a temperature
chaotic behavior for the system sizes considered.  We discuss the
experimental and theoretical implications of these findings.
\end{abstract}


\begin{multicols}{2}
\narrowtext
\parskip=0cm

The problem of chaos in spin glasses has been under
investigations for many years
\cite{PARISI,BRAYMOORE,BINDERYOUNG,KONDOR,NEYHIL,KONVEG,RITORT,FRANEY,NEYNIF}.  Even in the Sherrington-Kirkpatrick (SK) model, 
which is well understood with 
Parisi solution of the mean field theory\cite{MEPAVI},
the possible presence or absence of temperature chaos is still
an open problem. On the contrary, for
example, chaos induced by  a magnetic field $h$ was
already discussed by Parisi $15$ years ago \cite{PARISI}, and it is a
clear feature of the Replica Symmetry Breaking (RSB) scenario.  We
will give here numerical evidences of the fact that, for all lattice
sizes we are able to investigate by using state of the art optimized
Monte Carlo method \cite{MARINA}, there is no trace of temperature
chaos in mean field (infinite range) and realistic spin glasses,
in contradiction with previous 
claims\cite{BRAYMOORE,NEYHIL,KONVEG,RITORT,FRANEY,NEYNIF}. 
The question about
temperature chaos can be phrased by considering a typical equilibrium
configuration at temperature $T$, and one (under the same realization
of the quenched disorder) at $T'=T+dT$, where $dT$ is small: how
similar are such two configurations? In a chaos scenario for
any non-zero $dT$ the typical overlap would be exponentially small in the
system size.  We study both SK and  Diluted Mean
Field (DMF) \cite{DMF} models. 
We consider the DMF model in its version with constant
connectivity $c=6$.
Each lattice site is connected to $c$ other sites
chosen at random. It is interesting to check if this model has the
same features as the SK model. We also study the $3D$ Edwards Anderson
(EA) realistic spin glass. In all models spin variables are Ising like
($\sigma=\pm 1$), and the couplings $J$ can take the two values $\pm
1$ with probability $\frac12$. 
Our Monte
Carlo dynamics is based on {\em Parallel Tempering} (PT)
\cite{MARINA}:
we run in parallel two sets of copies of the system, and
always take overlap of configurations from two different Markov
chains. We use all standard precautions for checking thermalization of
our data \cite{MARINA}.  The indicator of a potential chaotic behavior
will be the two temperature overlap 
$q^{(2),(N)}_{T',T} \equiv
\overline{
\langle
\left( 
\frac1{N}
\sum_{i=1}^{N}
\sigma_i^{(T)}\tau_i^{(T')}
\right)^2
\rangle
}$.
The usual square overlap $q^{(2),(N)}_{T,T}$ is a
special case of $q^{(2),(N)}_{T',T}$.

\begin{figure}
  \centering
  \includegraphics[width=0.33\textwidth,angle=270]{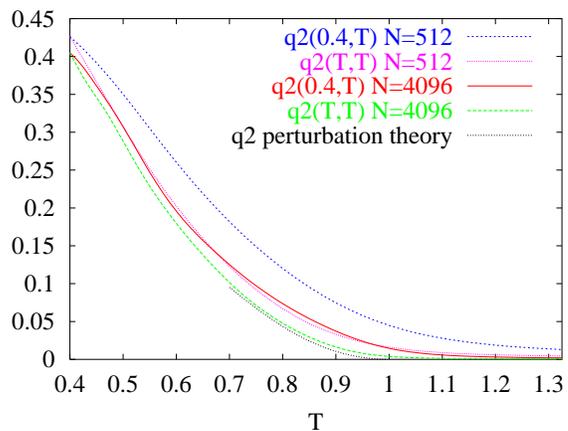}
  \caption[a]{
    $q^{(2)}$ at equal and different $T$ values for the SK model, 
    with $N=512$ and
    $N=4096$ sites.  The lower curve is the perturbative result for equal
    $T$ $q^{(2)}$.  See the text for details.}
  \protect\label{F-SKQ2Q2}
\end{figure}

Let us start from the analysis of our data for the SK model.  In
figure \ref{F-SKQ2Q2} we plot the square overlap for the two
temperature values $(0.4,T)$ (i.e. the overlap of a copy of the system
at temperature $T'=0.4$ with a copy of the system at $T\in
(0.4,1.35)$), and the one at equal temperature $(T,T)$.
The two upper (on the
left side of the plot) dashed curves (merging at $T=0.4$ at a value
close to $0.42$) are for $N=512$ spins (a small lattice size), the upper one
being the $(0.4,T)$ curve and the lower one the $(T,T)$ one. The two
lower curves (merging at $T=0.4$ at a value close to $0.40$) are for
$N=4096$ (our largest lattice for the SK model): of these two lower
curves the solid, upper curve is for $(0.4,T)$, while the dashed lower
one is the $(T,T)$ $q^{(2)}$. The fifth curve from the top, that stops
down at $T=0.7$ is the perturbative result for
$q^{(2),(\infty)}_{T,T}$ \cite{SOMMER} (useful for checking our
numerics and the quality of the approach 
to the asymptotic large volume limit). Here we only plot data
from two lattice sizes, and do not show the
statistical errors 
that are small
enough not to affect any of the issue discussed here, but would make
the picture less readable. We show data for the lowest temperature we have been
able to thermalize, $T'=0.4$. The same
qualitative picture holds for larger $T'$ values ($T'<T_c$).

One notices at first glance from figure \ref{F-SKQ2Q2} that for both
$N$ values (and, as we will see, for all $N$ values and different
systems we have analyzed) $q^{(2),(N)}_{0.4,T}$ $\gsim$ $ q^{(2),(N)}_{T,T}$\ 
($T\ge 0.4$). This is what happens in a non-chaotic systems 
(for example
ferromagnets where $q^{(2),(N)}_{T',T} = M(T)^2 M(T')^2$,
where $M(T)$ is the magnetization at temperature $T$), 
and is very different from what would happen in
a system with $T$-chaotic states. The second crucial observation is
that the distance between $q^{(2),(N)}_{0.4,T}$ and $q^{(2),(N)}_{T,T}$,
at fixed $T>0.4$, decreases with $N$, 
the two curves even seem to collapse at large $N$.
This kind of behavior shows the absence of temperature chaos in the
Sherrington-Kirkpatrick system and, as we will discuss in the
following, in the diluted mean field and  $3D$ EA
spin glasses. This evidence, together with an understanding of the
physical mechanism that is at the origin of this behavior (thanks to
the analysis of $P(q)$) is the main point of this note.

\begin{figure}
  \centering
  \includegraphics[width=0.33\textwidth,angle=270]{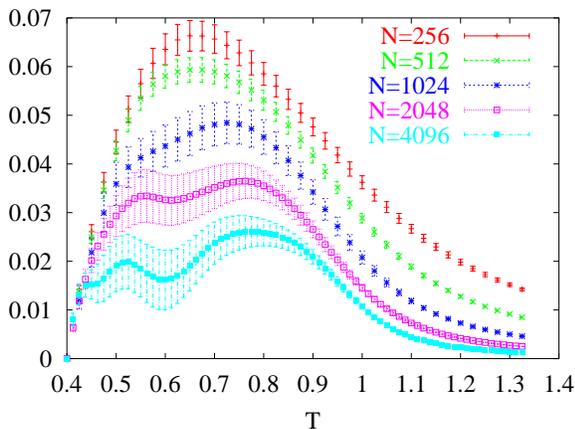}
  \caption[a]{
  $q^{(2),(N)}_{0.4,T} - q^{(2),(N)}_{T,T}$ as a function of $T$
  for the SK model with different $N$ values.
  }
  \protect\label{F-SKRATIO}
\end{figure}

A more quantitative evidence comes from figure \ref{F-SKRATIO}, where
we plot $q^{(2),(N)}_{0.4,T} - q^{(2),(N)}_{T,T}$ as a function of $T$
for the SK model with $N=256$, $512$, $1024$, $2048$ and $4096$.  Here
the errors are computed by an analysis of
sample to sample fluctuations  (it is important not to forget that
the points for different temperatures are strongly correlated, since
they involve the same $T=0.4$ data, or data from different temperatures
but nevertheless from the same PT simulation).  
In the large volume limit
both contributions to the difference are zero for $T>T_c=1$, so that the
non zero value of the curves in this regime gives us a measure of
finite size effects. Large lattices have larger
fluctuations. This is connected to the non-self-averageness of
$P_J(q)$: the peaks of $P_J(q)$ become narrower for large lattices
(eventually approaching $\delta$-functions in the large volume limit),
and averaging them to compute expectation values of the overlap gives
a wiggling behavior, that becomes smooth only for a very large
number of disorder samples.  We are not able to keep under control a
precise fit of the data of figure \ref{F-SKRATIO} for $N\to\infty$,
but the strong decrease of the difference of the data at large $N$ is
clear, and the possibility that the limit is zero everywhere looks
very plausible (it would be very interesting to understand 
theoretically this behavior).

\begin{figure}
  \centering 
  \includegraphics[width=0.33\textwidth,angle=270]{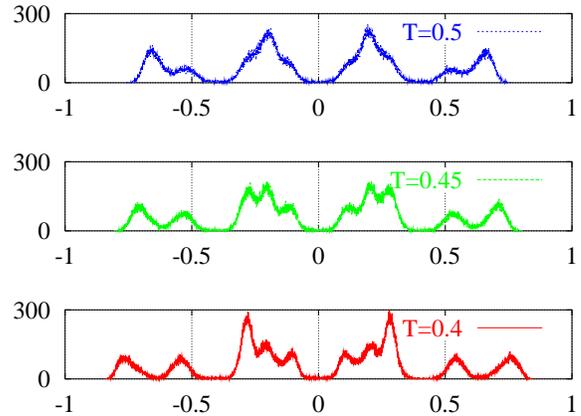}
  \caption[a]{ 
    $P_J(q)$ for the same selected disorder realization at three
    different temperatures of the
    SK model with $N=4096$ sites.} 
  \protect\label{F-SKPQ}
\end{figure}

We use figure \ref{F-SKPQ} for trying to understand better the
mechanism governing how stable states of the system vary as a function
of $T$. We plot the probability distribution $P_J(q)$ for a given
disorder realization of the SK model with $N=4096$. We show, from top
to bottom, the results at $T=0.50$, $0.45$ and $0.40$. 
The function  $P_J(q)$ should be symmetric around $q=0$, since we
are at zero magnetic field. The level of
symmetry reached by our finite statistics sample is a measure of 
the quality of our thermalization: from figure \ref{F-SKPQ} 
it looks very good. Note that there is no peak close to, or at, the
origin: this disorder realization carries little
weight in the $q\simeq 0$ region.
At the lowest $T$ value there are $5$ peaks for positive $q$, 
three of which very well separated. 
It is interesting to follow the
evolution of $P_J(q)$ from $T=0.50$ down to $T=0.40$. At $T=0.50$
there are basically two very broad peaks, that get resolved at
$T=0.45$: one broad peak gets divided in two clear peaks (that become
very clear at $T=0.4$), while the other forms a $3$ peak structure,
that get different weights at $T=0.4$. What one sees in figure
\ref{F-SKPQ} is interesting since it constitutes a typical pattern:
when lowering $T$, states start to contribute to the $P_J(q)$ by
bifurcations (new peaks emerge) and by smooth rearrangements of the
weights. One never sees dramatic changes involving strong
redistributions of weight among far away peaks, that would be typical
of a chaotic situation: the phase space is obviously very complex, as
it has to be in a situation characterized by RSB
\cite{MEPAVI}, but the $T$-dependence of the phase space is smooth and
non chaotic. 

\begin{figure}
  \centering 
  \includegraphics[width=0.33\textwidth,angle=270]{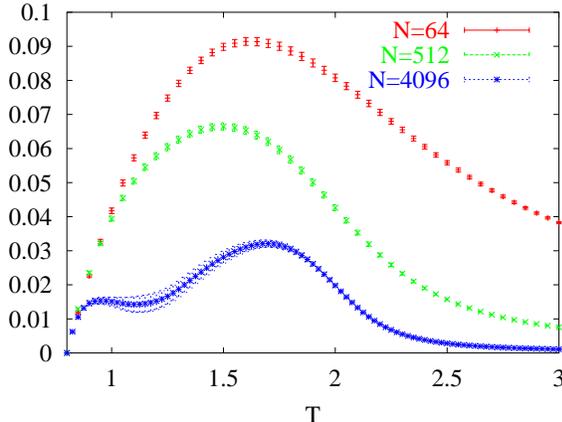}
  \caption[a]{ 
    As in figure \ref{F-SKRATIO}, but DMF, $N=64$, $512$ and $4096$.} 
  \protect\label{F-FI6RATIO}
\end{figure}

The situation in the DMF model (where $T_c\simeq 2.07$) is very
similar to the one in the SK model.  In figure \ref{F-FI6RATIO} we
show the analogous of figure \ref{F-SKRATIO}, for $N=64$, $512$ and
$4096$ spins. The two figures are very similar, and even the size of the
difference we are plotting is very similar in the two models, when
comparing the same values of $N$. The situation
in the DMF model looks exactly the same of the SK model: there is no
temperature chaos.

\begin{figure}
  \centering
  \includegraphics[width=0.33\textwidth,angle=270]{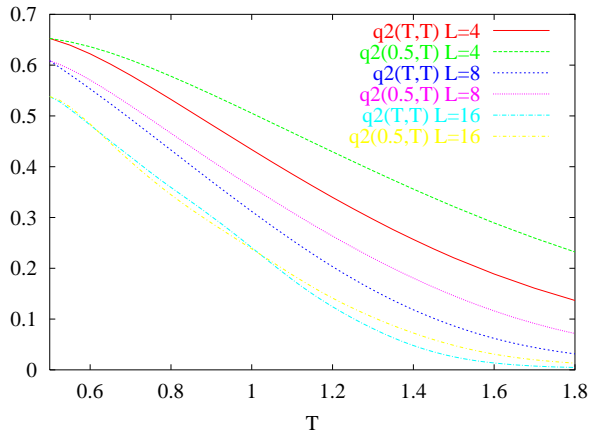}
  \caption[a]{
    $q^{(2)}$ at equal and different $T$ values for the $3D$ EA
  model.}
  \protect\label{F-3DQ2Q2}
\end{figure}

\begin{figure}
  \centering
  \includegraphics[width=0.33\textwidth,angle=270]{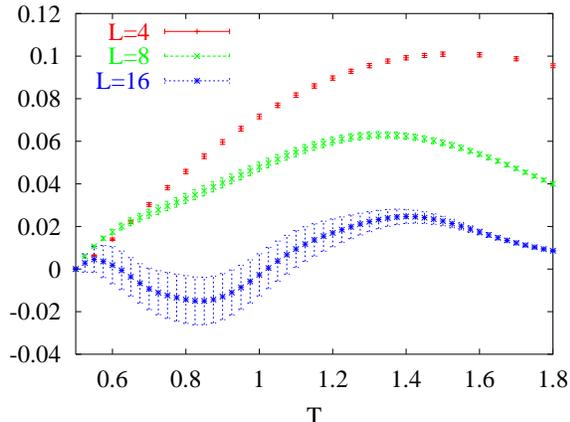}
  \caption[a]{
  $q^{(2),(N)}_{0.5,T} - q^{(2),(N)}_{T,T}$ as a function of $T$
  for the $3D$ EA model with different $L$ values.
  }
  \protect\label{F-3DRATIO}
\end{figure}

The situation in the $3D$ EA model is different only in that finite
size effects are very large (this is well known from numerical
simulations \cite{NUMERI}). In figure \ref{F-3DQ2Q2} we show $q^{(2)}$
at equal and different $T$ values for $L=4$, $8$ and $16$. It is clear
that $q^{(2),(N)}$ decreases noticeably with $N=L^3$ for all values of
$T$. It is also remarkable that even at very large $T$ values (with
$T$ far larger than the estimated value of 
$T_c\simeq 1.16$\cite{KAWYOU,CARPAL})
$q^{(2),(N)}$ is different from zero even at $N=4096=16^3$. Apart of
that figure \ref{F-3DQ2Q2} shows a situation very similar to the one
of \ref{F-SKQ2Q2}.
We are definitively not in a situation where 
$q^{(2),(N)}_{0.4,T}$ goes to zero exponentially and
$q^{(2),(N)}_{T,T}$ goes to a non zero limit
(even if the distance between the two curves for $L=16$ has
become very small, and even negative in a temperature region).  In
figure \ref{F-3DRATIO} we show the $3D$ analogous of figures
\ref{F-SKRATIO} and \ref{F-FI6RATIO}. Again, even for $T>T_c$, on the smaller lattices one has
non-zero differences: finite size effects are large, but apart from that
the emerging picture is analogous to the one we have found in mean
field (diluted and not).

Now, before discussing the data, we give a few details about our runs.
For the SK model we use $T_{min} = 0.4 = 0.4 T_c$.  We simulate
$N=256$, $512$, $1024$, $2048$ and $4096$: for the different $N$ cases
we have from $26$ to $142$ samples, a set of from $38$ to $75$
temperature values with a $dT$ going from $0.025$ to $0.0125$. We run
$200000$ sweeps but for the $N=4096$ and $N=2048$
lattices where we run $400000$
sweeps (we always use for measurement only the second part of the
run).  For the DMF model we have $T_{min} = 0.8 \simeq 0.4 T_c$.  We
use from $640$ to $1024$ samples for $N=64$, $512$ and $4096$.  Here
$T_{max}$ is $3$, the number of temperatures from $45$ to $89$ and the
number of iterations from $100000$ to $200000$.  In the $3D$ EA model
we use $T_{min}=0.4$, $T_{max}=2.075$
(here $T_c\simeq 1.16$) and a $dT$ going from $0.050$ to $0.025$.
We have $1344$ samples
for $L=4$ ($200000$ sweeps) and $L=8$ ($300000$ sweeps), and $64$
samples for $16^3$ (where $T_{min}=0.5$, with $3450000$ sweeps, this
is very many sweeps of many tempering copies).  Our SK program was
multi-spin coded on different sites of the same system (we store $64$
spins of the system in the same word), while the DMF and $3D$ codes
are multi-spin coded on different copies of the system \cite{MSC}.  We
want to note that, as compared to previous numerical simulations, we
have been able (thanks to a large computational effort and to the use
of PT) to thermalize the systems at very low $T$ values.  It is also
interesting to notice that in the $N=4096$ case the $3D$ EA model
requires many more sweeps than the DMF and the SK model.

In all our simulations we do not observe any temperature chaos effect.
This is true for the SK model, the diluted mean field and the $3D$ EA
model: the three models behave very similarly.  The differences we
have plotted, that would decrease exponentially in a chaotic scenario,
do not decrease faster than logarithmically.  Obviously from our
numerical findings we cannot be sure that things will not change for
very large system sizes, but again, we can claim that the absence of
any temperature chaotic behavior is crystal clear on our lattice
size. Two final comments are in order. At first, as we already said,
we cannot be sure about the behavior in the very large system limit:
the difference of $q^{(2)}_{T',T}$ and $q^{(2)}_{T,T}$ (for $T'\simeq
0.4 T_c$ and $T>T'$) decreases with the system volume, and is close to
zero on the larger lattice sizes we can simulate. This difference
could eventually become negative, and the correlation at $T\ne T'$
could eventually drop exponentially on very large lattices: we can
only say we do not see any trace of that. The second comment is that,
in any case, our results have an experimental relevance: the number of
spins that are equilibrated during a real experiment is of an order of
magnitude only slightly larger than the order of magnitude of the one
we can thermalize in our numerical simulations \cite{JOWHV}, so our
results strongly suggest the absence of temperature chaos in real
experiments.

The previous work of other authors on chaos was pointing toward the
presence of temperature chaos. On one side in this context the
analytic computations have by no means an unambiguous meaning, since
they are based on strong assumptions or on a perturbative and/or
approximate treatment. On the other side numerical computations of
older generations were much limited in scope as compared to what we
can do now. For example Ritort numerical computation \cite{RITORT},
that was correctly, in the limit of the gathered data, finding chaos,
was looking at a $T$ starting value of $0.4$, and a $dT$ of $0.5$ (as
compared to the $0.0125$ we have been able to use here), i.e. was
comparing $T=0.4$ to $T=0.9$ (where $T_c=1$) on reasonably small
lattice. In this case the decrease of the overlap is clear, but turns
out to be due to finite size effect (since even the equal $T$ overlap
has to go to zero at $T_c$).

A last comment (following for example \cite{AGING}) is about the
relevance of the absence of chaos for the description of realistic,
finite dimensional spin glasses.  In short the absence of a
temperature chaotic behavior makes impossible a modified droplet like
description of realistic spin glasses (the original droplet model
cannot work for example because of the observed dynamical scaling of
the energy barriers).

Following \cite{AGING} one notices that the very weak dependence of
spin glass physical properties on the cooling rate is not plausible in
a scenario of activated domain growth. Only arguing that there is
temperature chaos one can reconciliate the negligible effect of the
cooling rate with a droplet picture. The absence of temperature chaos
makes this reconciliation impossible.

We are aware that G. Parisi and T. Rizzo in a perturbative computation
close to $T_c$ find absence of temperature chaos (at the order they do
compute, but not necessarily at all orders in perturbation theory),
both in the SK and in the DMF model.  S.  Franz and I. Kondor have
connected evidence that excludes temperature chaos at lowest orders
close to $T_c$.  We deeply thank all of them, together with
J.-P. Bouchaud and F. Ritort, for interesting conversations. The
numerical simulations have used, together with a number of
workstations, computer time from the Grenoble T3E Cray and the Cagliari
Linux cluster Kalix2 (funded from Italian MURST under a COFIN grant).

\end{multicols}
\end{document}